# A MULTI-OBJECTIVE DIRECT ALGORITHM TOWARD STRUCTURAL DAMAGE IDENTIFICATION WITH LIMITED DYNAMIC RESPONSE INFORMATION

Pei Cao

Graduate Research Assistant and Ph.D. Candidate

Department of Mechanical Engineering

University of Connecticut

Storrs, CT 06269, USA

Qi Shuai

Research Scientist

Department of Mechanical Engineering

University of Connecticut

Storrs, CT 06269, USA

Jiong Tang$^{\uparrow}$

Professor

Department of Mechanical Engineering

University of Connecticut

Storrs, CT 06269, USA

Phone: (860) 486-5911, Email: jiong.tang@uconn.edu

---

$^{\uparrow}$ Corresponding author

# A Multi-Objective DIRECT Algorithm toward Structural Damage Identification with Limited Dynamic Response Information

Pei Cao, Qi Shuai and J. Tang

Department of Mechanical Engineering

University of Connecticut

Storrs, CT USA 06269

**ABSTRACT**

A major challenge in Structural Health Monitoring (SHM) is to accurately identify both the location and severity of damage using the dynamic response information acquired. While in theory the vibration-based and impedance-based methods may facilitate damage identification with the assistance of a credible baseline finite element model since the changes of stationary wave responses are used in these methods, the response information is generally limited and the measurements may be heterogeneous, making an inverse analysis using sensitivity matrix difficult. Aiming at fundamental advancement, in this research we cast the damage identification problem into an optimization problem where possible changes of finite element properties due to damage occurrence are treated as unknowns. We employ the multiple damage location assurance criterion (MDLAC), which characterizes the relation between measurements and predictions (under sampled elemental property changes), as the vector-form objective function. We then develop an enhanced, multi-objective version of the DIRECT approach to solve the optimization problem. The underlying idea of the multi-objective DIRECT approach is to branch and bound the unknown parametric space to converge to a set of optimal solutions. A new sampling scheme is established, which significantly increases the efficiency in minimizing the error between measurements and predictions. The enhanced DIRECT algorithm is particularly suitable to solving for unknowns that are sparse, as in practical situations structural damage affect only a small number of finite elements. A number of test cases using vibration response information are executed to demonstrate the effectiveness of the new approach.

## 1. INTRODUCTION

Structural Health Monitoring (SHM) has received significant attention in mechanical, civil and aerospace engineering communities in recent decades, since unpredicted structural failure may lead to catastrophic consequences. The latest advent of many new transducer materials/devices and the explosive progress in microelectronics and computational capability has further expedited the development of SHM systems. A typical SHM scheme uses structural dynamic responses measured by sensors to interpret the health status of the structure monitored. The dynamic responses are either the results of excitations by actuators that are part of an SHM scheme, or induced by ambient condition or normal operation. Usually, the dynamic responses are based upon wave/vibratory motions that can propagate quite far away from excitation sources. As anomalies in the dynamic responses are used to infer damage occurrence, consequently, these SHM schemes can cover very large structural area/volume.

One popular class of methods is wave propagation-based, which uses the change of transient wave (e.g., Lamb wave) upon its passage through damage site to infer damage occurrence (Michaels and Michaels, 2007; Harley and Moura, 2014). While these methods may entertain high detection sensitivity as high-frequency waves can be excited and employed as information carrier, it is generally difficult to use the transient responses to identify damage accurately, especially to quantify the severity (Cawley and Simonetii, 2005). An equally popular and traditional class of methods is vibration-based, which utilizes the changes of natural frequencies/modes of structures as inputs (West and Mafi, 1984; Kim and Stubbs, 2003). These changes of modal properties can be extracted from modal testing through scheduled actuation and sensing, or from ambient responses. One advantage of these methods is that they may be readily realized using off-the-shelf hardware. A qualitatively similar class of methods is the piezoelectric impedance-based methods (Annamdas and Radhika, 2013; Shuai et al, 2017; Cao et al, 2017b). In these methods, a piezoelectric transducer embedded to the host structure is driven by a frequency-swept sinusoidal voltage, which excites local stationary oscillations. The local structural oscillations in turn affect the electrical response of the transducer. As such, the impedance of the structure is coupled with that of the piezoelectric transducer. The change of piezoelectric impedance measured with respect to that under the undamaged baseline can be used as damage indicator.

A major challenge to virtually all SHM schemes is to accurately identify both the location and severity of damage using the dynamic response information acquired. It is worth noting that a potentially significant advantage of vibration-based and impedance-based methods is that, since the damage effect is reflected in the changes of stationary vibratory responses in these methods, it is possible to facilitate damage identification through inverse finite element analysis. Indeed, when a credible baseline finite element model of the structure monitored is available, we can first develop forward analysis of modal responses (e.g., eigenvalue problem) or impedance responses (under frequency-swept harmonic excitations), and then formulate an inverse sensitivity analysis based on Taylor series expansion under the assumption that each finite element is susceptible of damage occurrence. Using the measured response changes as input, the inversion of the sensitivity matrix can in theory yields the solution of the elemental property changes. Those elements with non-zero property changes indicate both the location and severity of damage in the finite element model.  Two issues, however, arise, especially in practical situations.  First, while the number of finite elements is generally large which leads to a large number of unknowns in the aforementioned inverse analysis, the available measurement information is usually limited.  For example, only a small number of modal frequencies and their changes can be realistically measured, and only the impedance responses around resonant peaks are truly sensitive to damage occurrence.  Therefore the inverse formulation may easily become under-determined (Kim and Wang 2014).  Second, oftentimes the input information is heterogeneous, encompassing different types of physical parameters.  For example, an SHM system with combination of vibration sensors and impedance sensors will produce different types of measurements.  For another example, in terms of vibration-based approach, both natural frequencies and mode shapes can be measured and used to infer damage occurrence.  In these cases, although more information generally leads to more accurate damage identification results, integrating different types of measurements together into a unified inverse analysis formulation is not straightforward.  The inevitable measurement noise and modeling uncertainty further compound the difficulty in inverse analysis (Shuai et al, 2017).

Alternatively, the problem of identifying damage location/severity in a finite element based inverse analysis using stationary response change as input can be cast into a global optimization formulation.  In such a formulation, the possible property changes in all finite elements are the unknowns to be solved. We sample these unknowns, and the differences between measured response changes and the predicted response changes (under sampled elemental property changes) obtained on the forward finite element analysis are treated as objectives to be minimized.  Several stochastic optimization approaches have been attempted, including the genetic algorithms (Hao and Xia, 2002; Villalba and Laier, 2012), particle swarm optimization (Seyedpoor, 2012), simulated annealing (He and Hwang, 2006), and differential evolution algorithms (Seyedpoor et al, 2015).  A common type of objective functions used are derived from multiple damage location assurance criterion (MDLAC) introduced by Messina et al (1998), which quantifies the relation between actual damage effect and model prediction in the unknown damage parametric space.  One advantage of solving the damage identification problem via global optimization is that heterogeneous measurements can be easily handled by establishing multiple objectives or vector-form objective functions.  For example, a set of MDLAC values, each corresponding to a type of measurements, can be employed to quantify the respective difference/error.  We can then resort to multi-objective optimization algorithms to solve the problem by minimizing such differences.

While the idea of employing global optimization to identify damage location and severity in a finite element model appears to be promising, the algorithms employed so far lack the efficiency and accuracy (Seyedpoor et al, 2015; Cao et al, 2017a).  For example, the number of possible damage locations is quite large because each element is susceptible of damage occurrence, and the severity level at each element is a continuous variable in theory. Thus a very large number of runs are required for stochastic optimization. Moreover, the results of stochastic techniques vary for each independent run with or without uncertainties. The performance is also dependent upon the algorithmic parameters that are usually tuned empirically. To tackle these issues and to make fundamental advancement in damage identification using limited dynamic response information, in this research we develop a multi-objective variant of a deterministic global optimization that follows the underlying idea of Dividing RECTangles, known as DIRECT (Jones et al, 1993). As the outcome of a deterministic technique, the results obtained are repeatable and conclusive without uncertainties. Only one algorithmic parameter, the number of function evaluations, is needed here. Specifically, a new sampling/division scheme in the unknown parametric space is established which significantly increases the efficiency in minimizing the difference between measurements and predictions. The enhanced DIRECT algorithm is particularly suitable to solving for unknowns that are sparse, as in practical situations structural damage affect only a small number of finite elements. The rest of the paper is organized as follows. In Section 2, we briefly outline the basic equations and optimization objectives involved in vibration-based damage identification which is used without loss of generality to illustrate the new method. In Section 3, the details of the proposed multi-objective DIRECT algorithm are presented to solve the damage identification problem. The proposed method is evaluated and validated for five benchmark damage scenarios in Section 4, where a comparative study on different strategies of the multi-objective DIRECT algorithm is also conducted. Concluding remarks are given in Section 5.

## 2. PROBLEM FORMULATION

Our objective here is to develop an efficient global optimization approach for damage identification. We focus on cases where the measured response information is limited and heterogeneous measurements are available. Without loss of generality we use the well-known vibration-based approach to illustrate the methodology formulation. Natural frequencies especially lower-order ones, albeit widely used for damage identification, are global properties that may not be sensitive to local damage. In certain cases mode shapes can be acquired which may be able to reflect better local effect of damage (Fan and Qiao, 2011). Combining these two sets of measurements can certainly yield better damage identification results.

Assume a credible finite element model of the structure monitored is available. The stiffness matrix of a healthy structure is denoted as $\mathbf{K}^R = \sum_{i=1}^{n} \mathbf{K}_i^R$, where $n$ is the number of elements, and $\mathbf{K}_i^R$ is the reference (healthy) stiffness of the $i$-th element. We assume damage causes stiffness change. The stiffness matrix of the structure with damage is denoted as $\mathbf{K}^D = \sum_{i=1}^{n} \mathbf{K}_i^D$, where $\mathbf{K}_i^D = (1-\alpha_i)\mathbf{K}_i^R$. $\alpha_i \in [0, 1]$ ($i=1,\cdots,N$) is the damage index for the $i$-th element. For example, if the $i$-th element suffers from damage that leads to a 20% of stiffness loss, then $\alpha_i = 0.2$. We further assume damping is negligible, and the modes are ortho-normalized. The $j$-th eigenvalue (square of natural frequency) and the $j$-th mode (eigenvector) are related as $\lambda_j = \{\phi_j\}^T \mathbf{K} \{\phi_j\}$. The change of the $j$-th eigenvalue from the healthy status to the damaged status can be derived as

$$\Delta\lambda_j = \{\phi_j\}^T (\mathbf{K}^D - \mathbf{K}^R)\{\phi_j\} = \sum_{i=1}^{n} \alpha_i \{\phi_j\}^T \mathbf{K}_i^R \{\phi_j\} \tag{1}$$

which can be written as

$$\Delta\lambda_j = \sum_{i=1}^{n} \alpha_i \cdot S_{ji} \tag{2}$$

or, in matrix/vector form,

$$\mathbf{\Delta\lambda} = \mathbf{S\alpha} \tag{3}$$

where $\mathbf{S}$ is the sensitivity matrix whose elements are given in Equation (1), and $\mathbf{\Delta\lambda}$ and $\mathbf{\alpha}$ are, respectively, the $q$-dimensional natural frequency change vector (based on measurement and baseline healthy results) and the $n$-dimensional damage index vector. Symbolically, the damage index vector can be expressed as

$$\mathbf{\alpha} = \mathbf{S}^{-1}\mathbf{\Delta\lambda} \tag{4}$$

It is worth noting that Equation (4) is usually underdetermined in engineering practices, because $n$, the number of unknowns (i.e., the number of finite elements), is much greater than $q$, the number of natural frequencies that can be realistically measured. This is one major reason that we want to avoid matrix inversion of $\mathbf{S}$ and resort to optimization (to minimize the difference between the measurements and predictions obtained from a model with sampled damage index values). Messina et al (1998) proposed a correlation coefficient, referred to as the multiple damage location assurance criterion (MDLAC), to compare two natural frequency change vectors, i.e., measured frequency change $\mathbf{\Delta\lambda}$ and predicted frequency change $\delta\mathbf{\lambda}$ (obtained based on assumed damage index values), as expressed below,

$$\text{MDLAC}(\mathbf{\Delta\lambda}, \mathbf{\alpha}) = \frac{\langle \mathbf{\Delta\lambda}, \delta\mathbf{\lambda}(\mathbf{\alpha}) \rangle^2}{\langle \mathbf{\Delta\lambda}, \mathbf{\Delta\lambda} \rangle \cdot \langle \delta\mathbf{\lambda}(\mathbf{\alpha}), \delta\mathbf{\lambda}(\mathbf{\alpha}) \rangle} \tag{5}$$

where $\langle *, * \rangle$ calculates the inner product of two vectors. $\text{MDLAC}(\mathbf{\Delta\lambda}, \mathbf{\alpha}) \in [0,1]$ captures the similarity between $\mathbf{\Delta\lambda}$ and $\delta\mathbf{\lambda}$. If the value of MDLAC equals to one, then these two vectors compared are identical in terms of direction. It has been shown that using MDLAC of natural frequencies as objective is sensitive to the damaged elements but may also find a healthy element as a damaged one (Nobahari and Seyedpoor, 2011). Thus, in order to take more local damage information into consideration, for the *j*-th mode shape, we can compare the measured change and predicted change using MDLAC in a similar manner,

$$\text{MDLAC}(\{\Delta\phi_j\}, \boldsymbol{\alpha}) = \frac{\left\langle \{\Delta\phi_j\}, \{\delta\phi_j(\boldsymbol{\alpha})\} \right\rangle^2}{\left\langle \{\Delta\phi_j\}, \{\Delta\phi_j\} \right\rangle \cdot \left\langle \{\delta\phi_j(\boldsymbol{\alpha})\}, \{\delta\phi_j(\boldsymbol{\alpha})\} \right\rangle} \tag{6}$$

Although multiple mode shapes may be measured in practice and incorporated into damage identification, to simplify presentation here we consider the case where only one mode, i.e., the *j*-th mode, is measured. Since both MDLACs are functions of $\boldsymbol{\alpha}$, and the measured data $\Delta\boldsymbol{\lambda}$ and $\{\Delta\phi_j\}$ are known, a multi-objective minimization problem for an *n*-element structure can then be formulated,

$$\text{Find: } \boldsymbol{\alpha} = \{\alpha_1, \alpha_2, ..., \alpha_n\}$$

$$\text{Minimize: } f_1 = -\text{MDLAC}(\Delta\boldsymbol{\lambda}, \boldsymbol{\alpha}), \ f_2 = -\text{MDLAC}(\{\Delta\phi_j\}, \boldsymbol{\alpha}) \tag{7}$$

$$\alpha^l \leq \alpha_i \leq \alpha^u, \ i = 1, 2, ..., n$$

where $\alpha^l$ and $\alpha^u$ are the lower bound and upper bound of the damage index. If more modes can be measured and are to be used in damage identification, we only need to add more objective functions into the optimization problem. The optimization problem defined is non-convex and may have many local optima. It is only appropriate when a global optimization approach is adopted.

## 3. MULTI-OBJECTIVE DIRECT ALGORITHM
### 3.1 DIRECT algorithm

DIRECT (Dividing RECTangles) algorithm is a deterministic global optimization algorithm evolved as a technique to address the shortcomings of the Lipshitzian optimization (Jones et al, 1993). It generally features higher efficiency than stochastic global techniques due to its deterministic attribute, and the optimization result is reproducible for each independent run without uncertainties. Furthermore, parameter tuning is not a necessity in DIRECT unlike most other global optimization algorithms. Such features are desirable when tackling damage identification problems because a conclusive result independent of the parameters selection can be expected.

To elaborate the DIRECT algorithm, we first provide the definition of Lipschitz continuity. Let $M \subset R^N$ and $f: M \to R$. The function $f(x)$ is said to be Lipschitz continuous on $M$ with Lipschitz constant $\gamma$ if we have:

$$|f(x) - f(x')| \leq \gamma \|x - x'\| \ \forall x, x' \in M \tag{8}$$

Then if function $f(x)$ is Lipschitz continuous on $[a,b]$ with constant $\gamma$, by replacing $x'$ in Equation (8) with $a$ and $b$ we can have the following

$$f(x) \geq f(a) - \gamma(x-a) \quad \forall x \in [a,b]$$
$$f(x) \geq f(b) + \gamma(x-b) \quad \forall x \in [a,b] \tag{9a,b}$$

As shown in Figure 1, the point of intersection for the two lines leads to the 1st estimation of the minimum of $f(x)$. Such discretization can be carried on to gradually approximate the minimum of $f(x)$ on $[a,b]$ (Shubert, 1972) as illustrated in Figure 2.

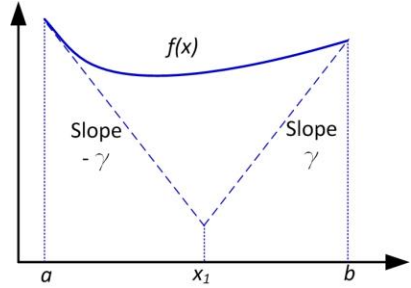

**Figure 1 Initial minimum estimation of $f(x)$ by Lipschitz optimization**

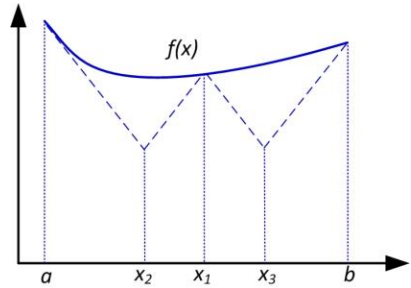

**Figure 2 The Shubert's algorithm**

However, there are several drawbacks of using Lipschitz optimization algorithms. First, they cannot be directly applied to high dimensional problems since the idea of endpoints does not exist when $N > 1$. Second, the Lipschitz constant $\gamma$ is usually hard to acquire. Subsequently, the original version of DIRECT algorithm was proposed to overcome these drawbacks (Jones et al, 1993; Finkel, 2005). Instead of using endpoints, DIRECT samples at midpoints of the parametric space which addresses the high dimensional challenge. Moreover, DIRECT relaxes the premise of $f(x)$ being Lipschitz continuous so there is no need for Lipschitz constant. To start the shift from Lipschitz optimization to DIRECT, we first assume that the Lipschitz constant is known and replace $x'$ in Equation (8) with $c = \dfrac{a+b}{2}$:

$$f(x) \geq f(c) + \gamma(x-c) \quad \forall x \in [a,c]$$
$$f(x) \geq f(c) - \gamma(x-c) \quad \forall x \in [c,b] \tag{10a, b}$$

These two inequalities correspond to the lines with slopes $\gamma$ and $-\gamma$ as demonstrated in Figure 3. Thus, the first estimation of the minimum with respect to $f(x)$ is obtained as $f(c) - \frac{\gamma(b-a)}{2}$, which only uses the function value at the center $f(c)$. Hence, the first problem of Lipschitz optimization regarding the tractability in high dimension has been solved. However, Lipschitz continuity is still assumed to hold and Lipschitz constant is considered as known. To complete the conversion, we specify the subdivision routine of DIRECT, i.e., where to sample and evaluate next. As shown in Figure 4, for interval $[a,b]$, the midpoint $c_1$ is sampled first, then the interval is divided into three segments and the function value is evaluated at midpoints of the smaller intervals $c_2$ and $c_3$. After partitioning interval $[a,b]$ into smaller intervals, next step in DIRECT, we determine which smaller interval should be further sampled.

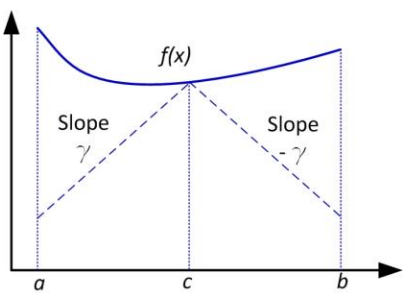

**Figure 3 Initial minimum estimation of $f(x)$ by center point sampling**

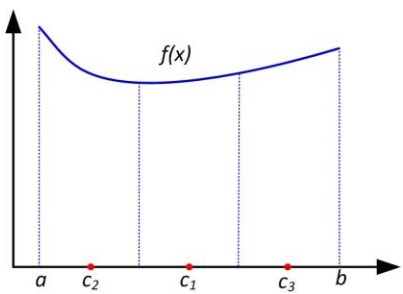

**Figure 4 Subdivision routine of DIRECT**

Let us assume that the parametric space has been partitioned into $m$ intervals $[a_i, b_i], i = 1,...,m$. Each interval can be represented by a point in a plane, as illustrated in Figure 5(a), where the horizontal axis stands for the distance from midpoint to vertices $\frac{b_i - a_i}{2}$, and the vertical axis is the function value $f(c_i)$. The fundamental idea of DIRECT is to determine which interval to be further sampled using the inequalities given below,

$$f(c_j) - K\left[(b_j - a_j)/2\right] \leq f(c_i) - K\left[(b_i - a_i)/2\right] \text{ for } i = 1,...,m$$

$$f(c_j) - K\left[(b_j - a_j)/2\right] \leq f_{\min} - \varepsilon |f_{\min}| \tag{11a, b}$$

where $\varepsilon > 0$ is a positive constant, $f_{\min}$ is the current best objective function value. If there exists some rate-of-change constant $K > 0$, then the interval $j$ is said to be potentially optimal and will be further evaluated (Figure 5(b)). Once the interval has been identified as potentially optimal, the interval is subdivided following routine given in Figure 4.

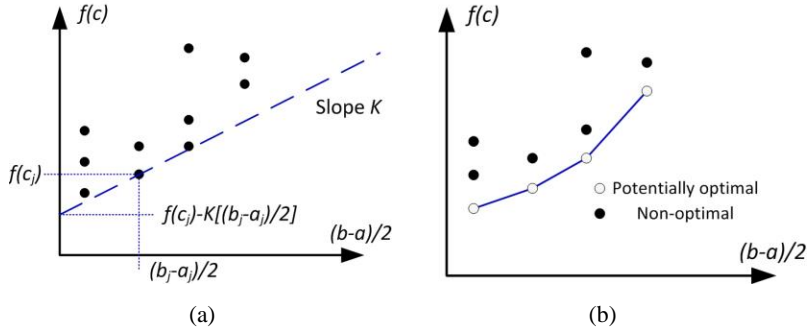

(a) (b)

**Figure 5 Determine the potentially optimal intervals in 1D**

The examples given so far are all in one dimension. Indeed, uni-variate DIRECT can be modified to multi-variate DIRECT by sampling the center points of hyper-rectangles. The multi-variate DIRECT algorithm can be summarized in four steps:

*1) Normalize*. First convert the decision domain to a unit cube/hyper-cube depending on the dimension of the decision space (Figure 6(a)).

$$\bar{\Omega} = \{x \in R^N : 0 \leq x_i \leq 1\}, \quad i = 1, 2, ..., n \tag{12}$$

*2) Sample and divide*. The center point $c_1$ of the cube/hyper-cube is sampled. Next, we evaluate the objective function values at

$$\mathbf{c}_1 \pm \omega \vec{e}_i, \quad i = 1, 2, ..., n \tag{13}$$

where $\omega$ is one-third of the length of the cube, and $\vec{e}_i$ is the $i$-th unit vector in Euclidean space (Figure 6(b)). The dimension $i$ will be divided first if $i$ satisfies,

$$\underset{i=1, 2, ..., n}{\arg\min} \left( \min(f(\mathbf{c}_1 + \omega \vec{e}_i), f(\mathbf{c}_1 - \omega \vec{e}_i)) \right) \tag{14}$$

Take Figure 6(c) for example. If $\min(f(\mathbf{c}_2), f(\mathbf{c}_3)) < \min(f(\mathbf{c}_4), f(\mathbf{c}_5))$, the dimension along $\mathbf{c}_2, \mathbf{c}_3$ direction will be divided first.

*3) Determine the potentially optimal rectangles*. This is the step that ensures the optimality of the solution. After step *1)* and step *2)*, the parametric space is divided into several rectangles/hyper-rectangles, e.g., five rectangles for 2D case (Figure 6(c)). A rectangle $j$ is said to be potentially optimal if there exists some rate-of-change constant $K > 0$ and positive constant $\varepsilon$ such that,

$$f(\mathbf{c}_j) - Kd_j \leq f(\mathbf{c}_i) - Kd_i, \quad \text{for } \forall i$$

$$f(\mathbf{c}_j) - Kd_j \leq f_{\min} - \varepsilon |f_{\min}| \tag{15a, b}$$

where $d_j$ measures the size of $j$-th rectangle. In the original version of DIRECT (Jones et al, 1993), $d$ is the distance from rectangle center $c$ to the vertex, and it is recommended to take $\varepsilon = 1 \times 10^{-4}$. Equation (15a) makes sure the rectangles selected are on the lower right of the collinear convex hull of the dots. Meanwhile, Equation (15b) prevents the algorithm from becoming too local in its orientation. Observe Figure 7. All the rectangles are projected to the plane where the horizontal axis denotes the rectangle size $d$ and the vertical axis represents objective value $f$. Each rectangle selected is the one with the smallest objective value in its $d$ class. And the selected rectangles are further divided and sampled by repeating step *2)*. In Figure 6(d), $\mathbf{c}_2$ and $\mathbf{c}_4$ are potentially optimal rectangles and will be therefore further divided.

*4) Repeat step 2) and 3) until maximum number of evaluation is reached.*

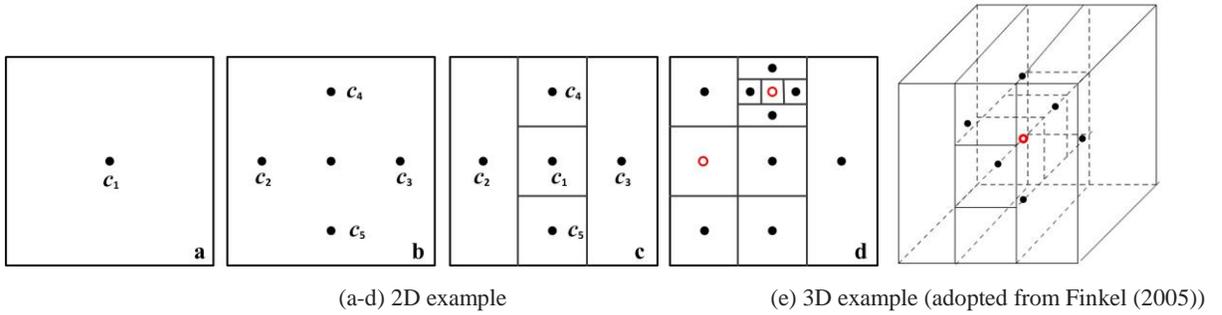

(a-d) 2D example  (e) 3D example (adopted from Finkel (2005))

**Figure 6 Sampling and dividing of the decision space**

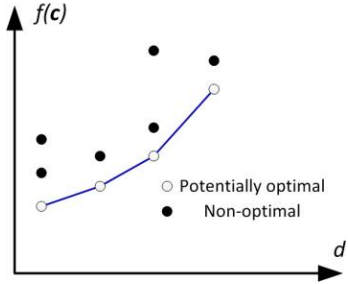

**Figure 7 Determine the potentially optimal rectangles**

In Figure 6, we provide 2D and 3D examples. Higher dimensional cases could be divided and sampled in a similar manner. The pseudo-code of DIRECT is given as follows.

Algorithm **DIRECT**

1: Normalize the search space to a unit cube/hypercube with center point $c_1$

2: Evaluate $f(\mathbf{c}_1)$; set $f_{\min} = f(\mathbf{c}_1)$, $\boldsymbol{\alpha}_{\min} = \mathbf{c}_1$, $k = 1$, and $k_{\max}$ = *max no. of evaluations*

3: **while** $k < k_{\max}$

4:     Identify the set **S** of potentially optimal rectangles following step *3)*

5:     **while** $\mathbf{S} \neq \varnothing$

6:         Take $j \in \mathbf{S}$

7:         Sample new points, evaluate *f* at the new points and divide the rectangle *j* following step *2)*

8:         Update $f_{\min}$, $\boldsymbol{\alpha}_{\min}$; $k := k + \Delta k$

9:         $\mathbf{S} := \mathbf{S} \setminus \{j\}$

10:    **end while**

11: **end while**

12: **Return** $\boldsymbol{\alpha}_{\min}$

### 3.2 Multi-objective DIRECT

The problem defined in Section 2 involves simultaneous optimization of 2 incommensurable and potentially conflicting objectives. Intuitively, multi-objective optimization (MOO) could be facilitated by forming an alternative problem with a composite objective function using weighted-sum approach. However, the weighted-sum methods have difficulties in selecting proper weighting factors. In order to provide the damage identification result independent of how the algorithm is implemented, *a posteriori* preference articulation is usually preferred, which allows a greater separation between the algorithm and the decision-making process (Cao et al, 2016). Accordingly, instead of a single optimum produced by weighted sum methods, MOO will generate a set of alternative solutions explicitly exhibiting the tradeoff between different objectives. Because of these advantages, the recent decade has seen a few attempts to extent DIRECT for multi-objective optimization problems. Wang et al (2008) proposed a hybrid algorithm which uses a multi-objective DIRECT algorithm to generate population for evolutionary algorithm. In their study, rank strategy is used to rank the rectangles into fronts in terms of their objective values, and then replace $f(\mathbf{c})$ in Equations (15a) and (15b) with the rank to identify the potentially optimal rectangles. More recently, Al-Dujaili and Suresh (2016) treated *d* as an additional objective value and obtain the potentially optimal rectangles by calculating the non-dominated Pareto front. Wong et al (2016) replace $f(\mathbf{c})$ in Figure 7 with the hypervolume indicator and select rectangles on the upper right Pareto front in the *hypervolume-d* plane.

In this research, we incorporate a new strategy into DIRECT technique to fulfill the multi-objective capability. When determining the potentially optimal rectangles, we first rank the rectangles into fronts/surfaces using non-dominated sorting (Deb et al, 2002) as illustrated in Figure 8, where two objectives are used in the example. The first front will be given a rank index *R*=1, the second front will have a rank index *R*=2, and so on so forth. Then the potentially optimal rectangles can be obtained by projecting the rectangles to *R-d* plane first and extract the lower right Pareto front as shown in Figure 4(a). In the case of *m*-objective minimization, a rectangle *j* is said to be potentially optimal if

$$R\big(R\big(f_1(c_j),\ f_2(c_j),\ldots,f_m(c_j)\big),-d\big)=1 \tag{16}$$

where *R(\*)* is an non-dominated sorting operator in minimization sense which returns the rank index. Recall that in the original implementation of DIRECT, *d* is the distance measured between the rectangle center and vertex. In this research, *d* is set to be the length of the longest side of the rectangle instead, which allows the algorithm to group more rectangles at the same size (Finkel, 2005). Compared to the original DIRECT algorithm, we omit the second condition (Equation (15b)) because by considering lower right Pareto front, the rectangles on the lower right convex hull that have the same *R* but smaller *d* will be automatically filtered out (Figure 9(a)), which resembles the effect of Equation (15b).

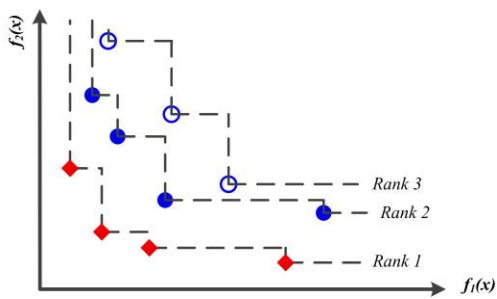

**Figure 8 Rank based on objective values**

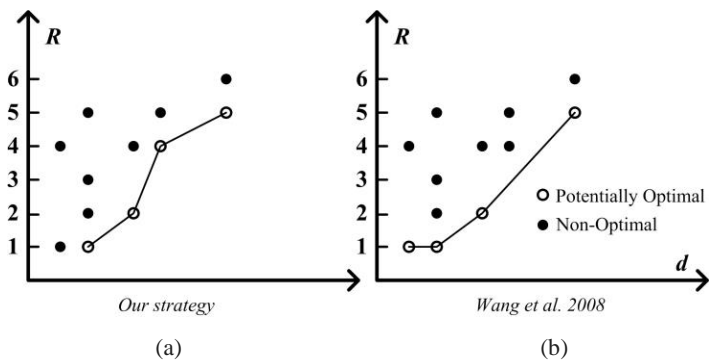

**Figure 9 Comparison of two strategies on selecting potentially optimal rectangles**

A major difference between proposed strategy and the strategy by Wang et al (2008) is that our strategy explores the concave rectangles as well (Figure 9). Generally, a low-ranked solution (large rank index) is dominated by some of the high-ranked (small rank index) solutions but meanwhile non-dominated to other high-ranked solutions. As a result, a portion of low-ranked rectangles may lead to optimal solutions just like certain high-ranked rectangles. Examining the lower right Pareto helps to partition some low-ranked rectangles with potential, thus gains an edge over the NS-DIRECT (Wang et al, 2008) in terms of exploration. Moreover, the proposed method overcomes the drawback of the MO-DIRECT (Dhjaili and Suresh, 2016), as we concentrate only on a small portion of the rectangles. MO-DIRECT, on the contrary, incorporates *d* as an additional objective which causes most of the rectangles fall into the potentially optimal group. Lastly, when competing with MO-DIRECT-hv (Wong et al, 2016), the proposed method is more efficient computationally. In MO-DRIECT-hv, hypervolume need to be calculated iteratively, whereas hypervolume calculation is a computationally expensive NP-hard problem. In Section 4, we will evaluate the performance of proposed algorithm given several damage cases and carry out comparison between the four-above-mentioned different multi-objective DIRECT techniques.

### 3.3 *A posterior* articulation

In optimization, when a decision maker expresses a subjective judgment before or during the optimization, a single biased solution is obtained. However, in engineering practice, it is quite common that several solutions are of interests. As mentioned in Section 3.2, it is preferable that the problem-solving and decision-making processes are independent of each other so that the problem-solving approach could remain unchanged when user judgments differ. Thus, multi-objective DIRECT optimization is adopted in this research, which will provide a set of solutions as damage identification candidates. Nevertheless, to obtain a single satisfactory candidate that compromises between objectives, *a posterior* articulation should be incorporated into the formulation as well.

Usually, damage index vector $\boldsymbol{\alpha}$ is sparse by nature since normally only a small portion of the structure monitored is affected by damage occurrence. In this research, we use a joint $l_0$ and $l_1$-norm as a sparse filter to pinpoint one sparse candidate in the solution set.

$$p = \arg\min_{i \in \mathbf{S}_e}(\|\boldsymbol{\alpha}_i\|_0 + \|\boldsymbol{\alpha}_i\|_1) \tag{17}$$

where $i$ is the index of the solution in optimal set $\mathbf{S}_e$, and $p$ is the index of the solution selected. In Equation (17), $l_0$-norm emphasizes on the number of damages, while $l_1$-norm emphasizes on finding the solution with the least amplitude. Consequently, a concise and interpretable sparse solution will be selected in this study as the posterior damage identification result. Since *a posterior* articulation is independent of the algorithm formulation, it could be modified at all times according to users' needs.

### 4. CASE STUDIES

In this section, we employ the proposed multi-objective DIRECT algorithm to five damage identification cases, as explained in Table 1. To facilitate easy re-production of case analyses, a benchmark cantilever beam model with varying number of elements and varying damage scenarios is adopted here for demonstration purpose. Case 1 and Case 2 serve as baseline scenarios. Case 3, Case 4 and Case 5 have increased difficulty level in terms of damage identification, which are employed to further investigate and interrogate the factors that may affect the performance and the outcome of the new algorithm.  Algorithm comparisons are carried out subsequently.

The Young's modulus of the beam is 69 GPa, the length per element is set as 10 m, and the area of cross-section is 1 m$^3$.  The measured mode shape and natural frequencies used are simulated directly from the finite element models which are subject to $\pm 1.5‰$ standard Gaussian uncertainties (Messina et al, 1998; Guo and Li, 2009; Seyedpoor et al, 2015).  Unless noted otherwise, the measurements available to the algorithm are limited to the first 5 natural frequencies and the 2$^{nd}$ mode shape for all cases.  It is worth mentioning here that the algorithm does not know the number of damages.  Nor does it know the locations and severities.  As discussed in Section 1, an important advantage of multi-objective DIRECT against other optimization techniques is that it has significantly less parameters to tune.  The only parameter needs to be specified here is the maximum number of evaluations, which is set to 30,000 as it is adequate for all cases to converge.  The average computational time for 30,000 function evaluations is 79 seconds within MATLAB on a 2.40GHz Xeon E5620 computer.

**Table 1 Numerical test case 1-5**

| Case | No. of elements | No. of damages | Location | Severity (stiffness loss) |
|---|---|---|---|---|
| **1** | 15 | 2 | 3; 14 | 9%; 5% |
| **2** | 15 | 2 | 8; 11 | 2%; 8% |
| **3** | 15 | 3 | 3; 9; 14 | 9%; 3%; 5% |
| **4** | 20 | 2 | 8; 11 | 2%; 8% |
| **5** | 30 | 3 | 8; 11; 21 | 2%; 8%; 4% |

**4.1 Case 1 (15 elements, 2 damages)**

We first carry out the case study on a 15-element cantilever beam.  The damages are on the 3$^{rd}$ and 14$^{th}$ element with severities $\alpha_3 = 0.09$ and $\alpha_{14} = 0.05$ respectively.  A set of optimal candidates are obtained (Figure 10(a)) as the result of tradeoff between objectives.  Each solution obtained corresponds to one possible damage situation.  Figure 10(b) shows the mean and variance of the damage indices for each element.  The mean value is then compared to the true damage indices in Figure 10(c).  As illustrated, by adopting the proposed approach, locations and severities of the damages are successfully identified.  The mean value of the identification results bears some small errors in element 5 and element 7.  By using *a posterior* articulation introduced in Section 3.3, the damages are better approximated (Figure 10(d)).  To better demonstrate the selection process of potentially optimal rectangles of the proposed method, several iterations of the multi-objective DIRECT search in objective space are presented (Figure 11), where each dot corresponds to one or a group of rectangles.  The vertical axis is the rank index *R* of the rectangles and horizontal axis is the size of the rectangle *d*.

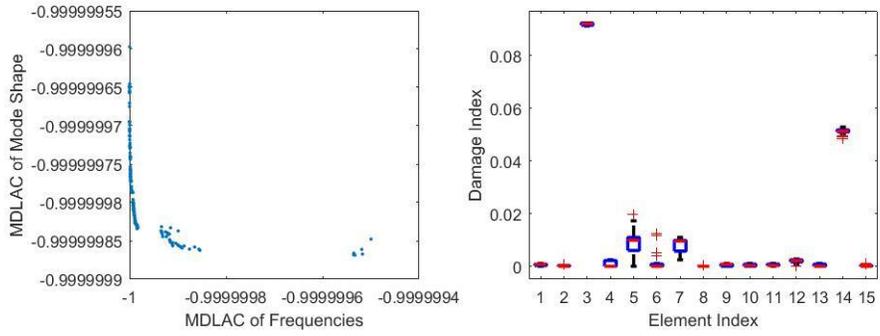

(a) In objective space  (b) Box plot

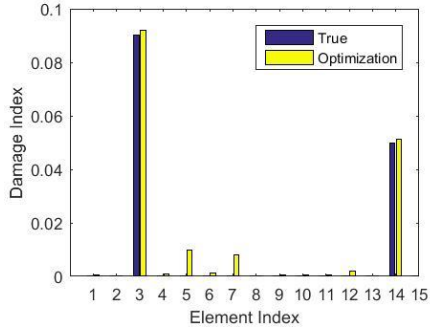 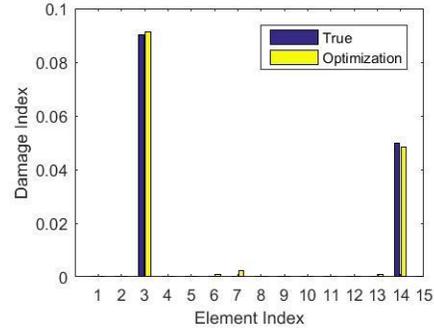

(c) Mean value  (d) After *a posterior* articulation

**Figure 10 Optimization results of Case 1**

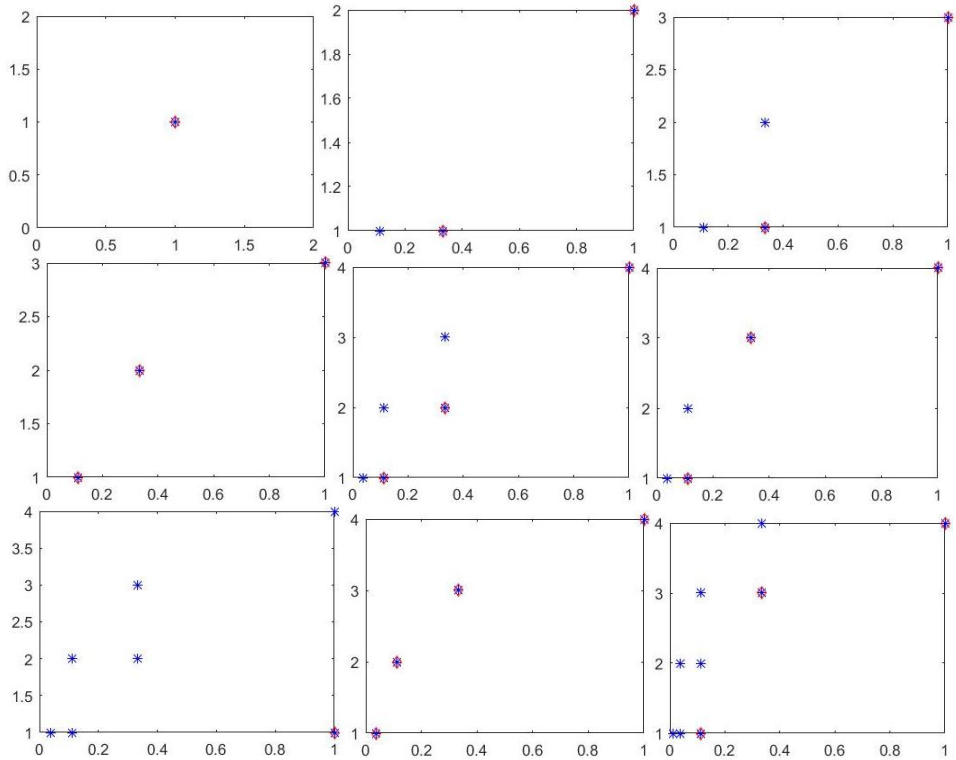

**Figure 11 Several iterations of multi-objective search in objective space (x-axis: *d*; y-axis: *R*)**

(◇ : **potentially optimal rectangles**; ∗: **rectangles**)

## 4.2 Case 2 (15 elements, 2 damages)

We then carry out another case study on the same 15-element cantilever beam. Different from Case 1, the damages are now on the $8^{th}$ and $11^{th}$ element with severities $\alpha_8 = 0.02$ and $\alpha_{11} = 0.08$ respectively. Together with Case 1, we try to investigate how different locations and severities of the damages affect the results while other factors remain unchanged. Similarly, a set of optimal candidates are obtained (Figure 12(a)), which could belong to the same solution niche due to small variance (Figure 12(b)). As illustrated in Figure 12(c), the locations and severities of the induced damages are accurately identified with only small errors, which are further reduced through *a posterior* articulation as depicted in Figure 12(d). For both Case 1 and Case 2, the proposed algorithm exhibits satisfactory performance regardless the locations and severities. Figure 13 gives the convergent history of DIRECT as applied to Case 2, where the horizontal axis is the number of function evaluations, and the vertical axis is the mean MDLAC value of the solutions in current optimal set. Unlike gradient-based optimization, both MDLAC curves fluctuate rather than monotonic decrease. This ensures the algorithm to be able to escape from local optima. Specifically, when the MDLAC curve ascends, new areas of the decision space are explored and sampled.

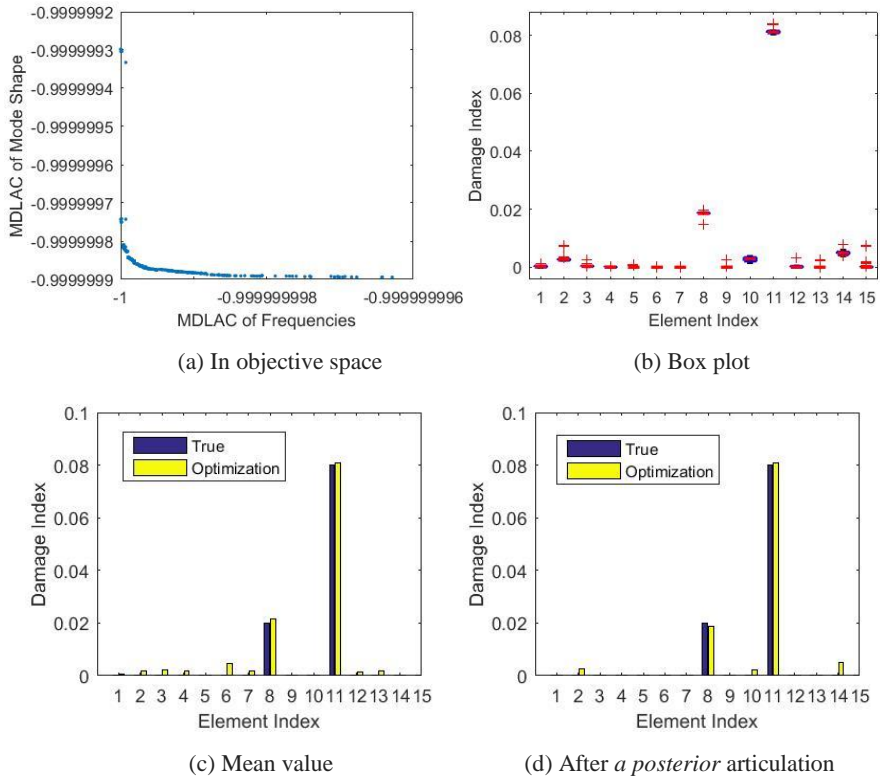

(a) In objective space      (b) Box plot

(c) Mean value      (d) After *a posterior* articulation

**Figure 12 Optimization results of Case 2**

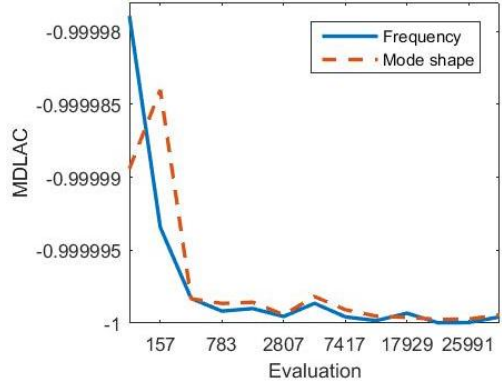

**Figure 13 Convergent history of Case 2**

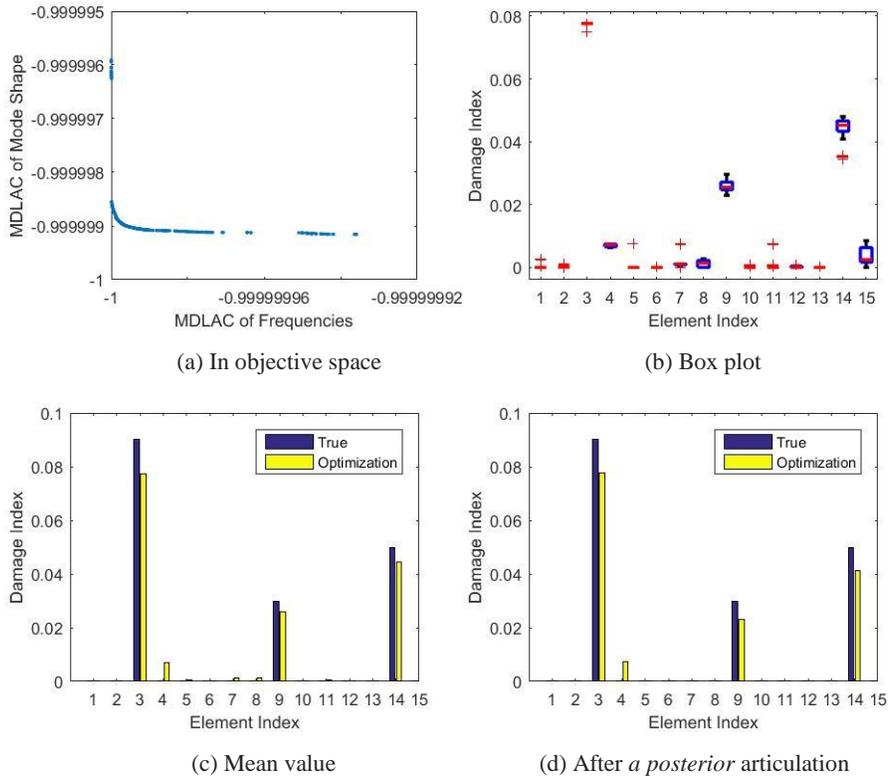

(a) In objective space      (b) Box plot

(c) Mean value      (d) After *a posterior* articulation

**Figure 14 Optimization results of Case 3**

## 4.3 Case 3 (15 elements, 3 damages)

Different from Case 1, in this Case 3 an additional damage is introduced to the $9^{th}$ element with severity of 0.03. The purpose of such a test is to examine the effect of the number of damages to damage identification accuracy. As can be seen from Figures 14(c) and 14(d), the locations of the damages are detected accurately. The results on damage severities are also very good, with only small discrepancies between actual damages and identified damages in terms of severity. Apparently, the optimal solutions fall into a "local minimum" of the original objective space because the current objective space is deformed by the measurement errors introduced. On the other hand, because of the center-sampling nature of the DIRECT algorithm, the more damages there are, the less sparse the damage index vector is, which adds extra burden to the algorithm computationally. It is shown in Figure 15 that it indeed takes the algorithm longer to converge than in Case 2.

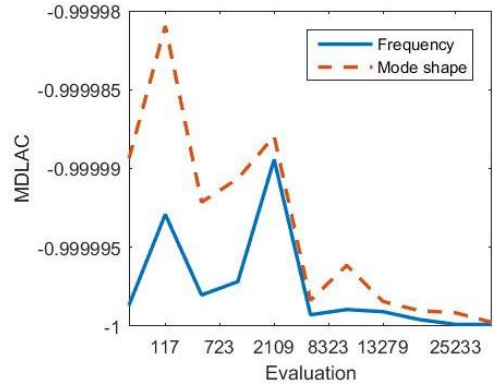

**Figure 15 Convergent history of Case 3**

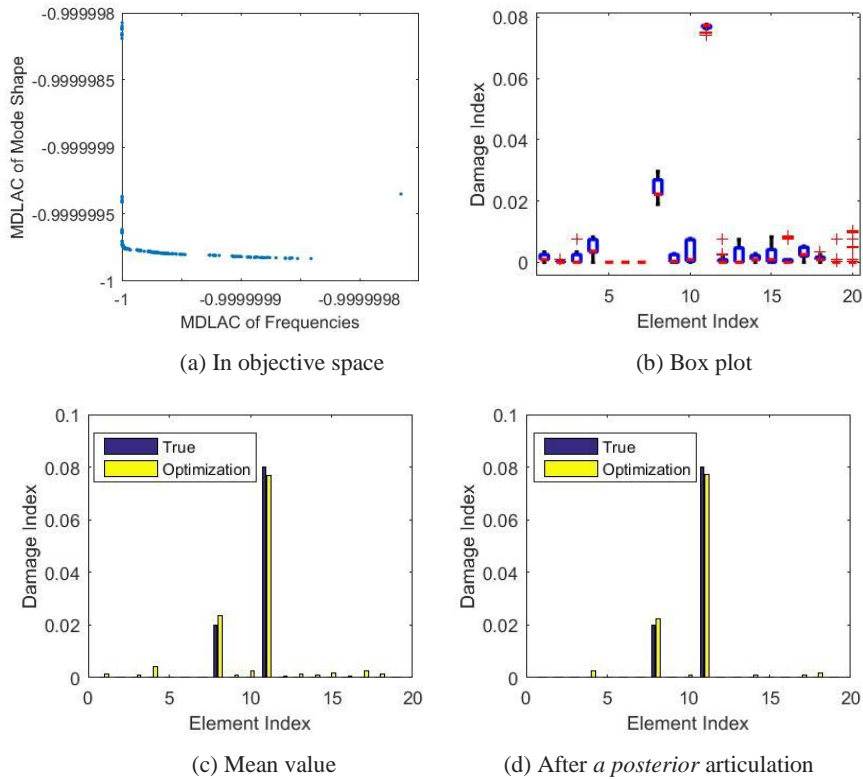

**Figure 16 Optimization results of Case 4**

### 4.4 Case 4 (20 elements, 2 damages)

Case 4 is derived from Case 2, where a 20-element beam is used rather than a 15-element beam. In optimization perspective, it is equivalent to adding 5 variables. The results on locations and severities are still generally good even with more variables. As shown in Fig 16(b), the solutions have larger variance compared to that of Case 2 (Figure 12(b)). Due to enlarged search space, the solutions also tend to be less clustered (Figure 16(a)). Nevertheless, the proposed algorithm locates the induced damages with very good accuracy (Figure 16(c)), and the errors are further eliminated by *a posterior* articulation proposed. For an ideal model without uncertainty, adding variables alone would not change the essence of the problem. In other words, if the optimization process lasts long enough, the quality of the final solutions would not deteriorate. However, errors and uncertainties are inevitable in engineering practices. Under realistic circumstances, more variables require not only more computational time, but also more measurements. Otherwise, the accuracy of the results expects to decrease. That is why the optimization results of Case 4 reported in Figure 16 are not as good as those of Case 2.

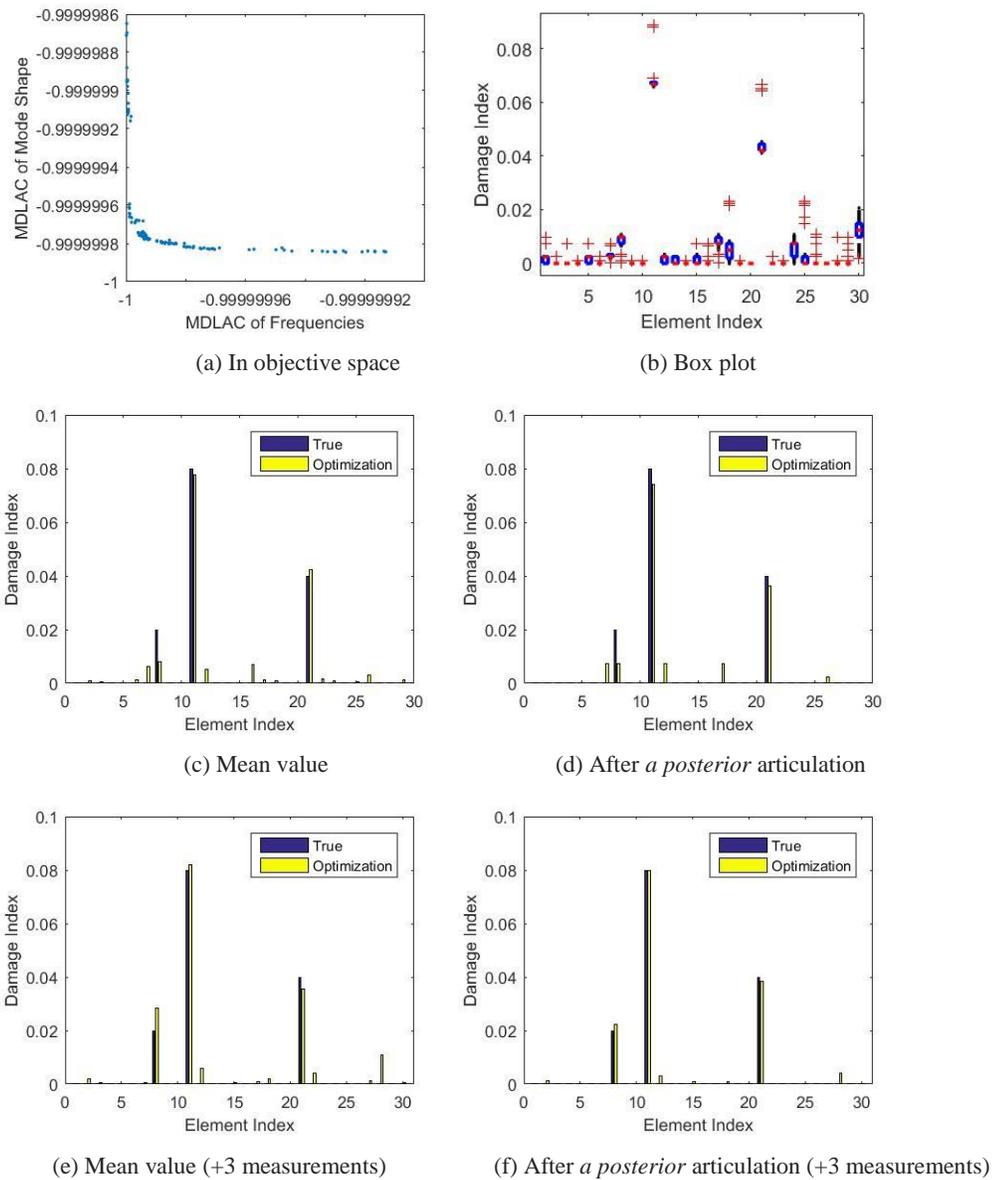

(a) In objective space    (b) Box plot

(c) Mean value    (d) After *a posterior* articulation

(e) Mean value (+3 measurements)    (f) After *a posterior* articulation (+3 measurements)

**Figure 17 Optimization results of Case 5**

### 4.5 Case 5 (30 elements, 3 damages)

In Case 5, 10 more elements and 1 more damage are added to the damage scenario given in Case 4 (Table 1). As have been discussed in Case 3 and Case 4, either more damages or more variables will affect the quality of the solution negatively. Therefore, in Case 5, the optimal solutions suffer the most error among all numerical test cases conducted in this study (Figure 17). The results, however, can still be considered acceptable. Moreover, when first 8 natural frequencies are used instead of 5, the identification results are improved considerably (Figures 17(e) and 17(f)) even if the problem is still ill-posed.

In summary, the proposed approach has accurately identified both locations and severities for all numerical test cases given limited and contaminated data.

### 4.6 Algorithm comparison

In this sub-section, we perform the comparison of the proposed method with other multi-objective DIRECT techniques mentioned in Section 3.2, namely NS-DIRECT, MO-DIRECT and MO-DIRECT-hv, on Case 3 (described in Section 4.3). The main differences between these algorithms are the strategies used when determining the potentially optimal rectangles. The strategy used in our proposed approach, NS-DIRECT, MO-DIRECT and MO-DIRECT-hv are denoted as Pareto front strategy, non-dominated strategy, convex hull strategy, and hypervolume strategy, respectively. Different from Section 4.3, for comparison purpose, first 9 natural frequencies are used, and the uncertainties are omitted. Table 2 records the predicted results compared to true damage indices for the 3 damaged elements, in which the best predictions are marked in gray. All predictions made by the proposed algorithm are ranked the best among the four algorithms. The performance of the algorithm is also depicted in Figure 19 as a spider plot. Each vertex of the polygon represents the error between the true damage index for the specific element. As can be seen in Figure 19, among all the strategies, the black polyline, which represents the Pareto front strategy used in our proposed approach, has the best overall performance.

**Table 2 Algorithm comparison**

| Element<br>Algorithm | True vs. predicted damage indices for damaged elements: mean value | | |
|---|---|---|---|
| | 3 | 9 | 14 |
| True | 0.09 | 0.03 | 0.05 |
| Proposed DIRECT | 0.0900 | 0.0300 | 0.0500 |
| NS-DIRECT | 0.0875 | 0.0425 | 0.0575 |
| MO-DIRECT | 0.0897 | 0.0302 | 0.0507 |
| MO-DIRECT-hv | 0.0859 | 0.0254 | 0.0409 |

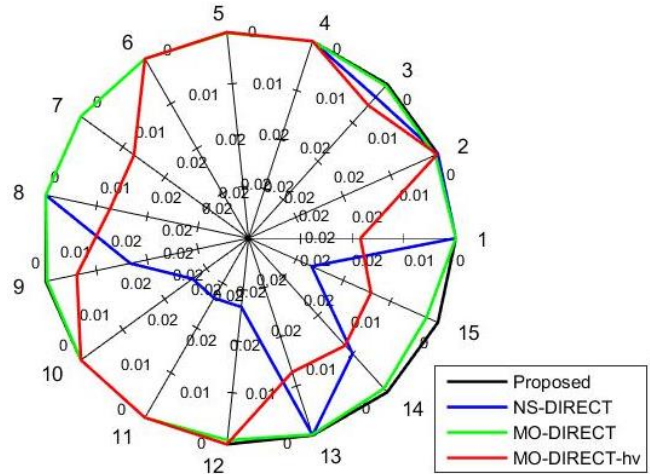

**Figure 18 Strategy comparison of multi-objective DIRECT algorithms**

## 5. CONCLUDING REMARKS

This research aims at solving the fundamental challenge in damage identification using limited dynamic response information and heterogeneous measurements. With a vibration-based approach as an illustrative example, we formulate a multi-objective optimization problem that can be solved to yield damage location and severity. An enhanced multi-objective DIRECT, which is a global deterministic optimization technique, is formulated. The algorithm has less algorithm parameters compared to stochastic optimization techniques. Due to its deterministic nature, the identification results are conclusive and repeatable. A new sampling scheme is established in this research which significantly increases the efficiency in minimizing the error between measurements and predictions. The enhanced DIRECT algorithm is particularly suitable to solving for unknowns that are sparse in nature since in practical situations structural damage affect only a small number of finite elements. Comprehensive case analyses are performed that demonstrate the effectiveness of the new method. This method can be applied to a variety of damage identification problems.

## ACKNOWLEDGMENT

This research is supported in part by AFOSR under grant FA9550-14-1-0384.